\documentclass[preprint,12pt]{elsarticle}
\usepackage{amssymb}
\usepackage{graphicx}
\usepackage{subfigure}
\usepackage{multirow}
\usepackage[ruled,vlined]{algorithm2e}
\usepackage{algorithm2e}

\journal{Parallel Computing}

\begin{document}

\begin{frontmatter}

    \title{A Critical Path Approach to Analyzing Parallelism of Algorithmic Variants. Application to
    Cholesky Inversion.}

    \author{Henricus Bouwmeester}
    \ead{henricus.bouwmeester@ucdenver.edu}
    \author{Julien Langou}
    \ead{julien.langou@ucdenver.edu}
    \ead[http://math.ucdenver.edu/~langou/]{http://math.ucdenver.edu/~langou/}
    \address{Dept. of Mathematical and Statistical Sciences 
    University of Colorado Denver}
    \address{Research was supported by the National Science Foundation grant \# NSF CCF-811520}

    \begin{abstract}

Algorithms come with multiple variants which are obtained by changing the
mathematical approach from which the algorithm is derived. These variants offer
a wide spectrum of performance when implemented on a multicore platform and we
seek to understand these differences in performances from a theoretical point
of view. To that aim, we derive and present the critical path lengths of each
algorithmic variant for our application problem which enables us to determine a
lower bound on the time to solution.  This metric provides an intuitive grasp
of the performance of a variant and we present numerical experiments to
validate the tightness of our lower bounds on practical applications.  Our case
study is the Cholesky inversion and its use in computing the inverse of a
symmetric positive definite matrix.  

    \end{abstract}


\end{frontmatter}

\section{Introduction}
\label{intro}

An algorithm can be decomposed into specific tasks which have dependencies on
other tasks such that a directed acyclic graph (DAG) can be formed by drawing
all of these tasks as nodes and the dependencies as edges between the nodes.
By doing so, the longest path of tasks from the first task(s) of the algorithm
to the final task(s) describes the critical path.  By changing the weights of
the tasks, the critical path (and its length) may also change accordingly.

Our study will involve so called {\em tiled algorithms} whose individual tasks
are part of BLAS and LAPACK and are executed sequentially by a core; all these
tasks are then scheduled dynamically on a multicore platforms.  Tiled
algorithms with a dynamic scheduler in the context of multicore architectures
have been presented in~\cite{Buttari2008,tileplasma,Quintana:2009} for the
Cholesky factorization, LU factorization and QR factorization.  This paradigm
is the idea behind the PLASMA software~\cite{plasma_users_guide}.  From 2008 to
2010, numerous papers have been written on presenting the
performance,
improving the scheduling, auto-tuning of these
algorithms, presenting new variants of these algorithms, and extending this
paradigm to others algorithms and to parallel architectures other than
multicore platforms.  In the context of the Cholesky inversion problem, (which
is the application subject of this paper,) the corresponding tiled algorithms
were presented in~\cite{Quintana:2009}.

The Cholesky inversion of a symmetric positive definite matrix will consist of
three steps:  Cholesky factorization, inversion of the Cholesky factor,
multiplication of the transpose of the inverse with itself. (See
Algorithm~\ref{alg:InPlace}.)

We first tile the $n \times n$ SPD matrix $A$ into $t \times t$ tiles of size
$b \times b$ and without loss of generality consider $n = t \cdot b$. We
consider here $t=4$. Then, the first step of the algorithm is TILE\_POTRF.
(See Algorithm~\ref{alg:InPlace} Step 1.)
It computes the Cholesky factor $L$ such that
\[ 
\left[ \begin{array}[h]{cccc}
    A_{11} & \cdot & \cdot & \cdot \\
    A_{21} & A_{22} & \cdot & \cdot \\
    A_{31} & A_{32} & A_{33} & \cdot \\
    A_{41} & A_{42} & A_{43} & A_{44} \\
\end{array}\right]
=
\left[ \begin{array}[h]{cccc}
    L_{11} &   &   &   \\
    L_{21} & L_{22} &   &   \\
    L_{31} & L_{32} & L_{33} &   \\
    L_{41} & L_{42} & L_{43} & L_{44} \\
\end{array}\right]
\left[ \begin{array}[h]{cccc}
    L_{11}^{T} & L_{21}^{T} & L_{31}^{T} & L_{41}^{T} \\
      & L_{22}^{T} & L_{32}^{T} & L_{42}^{T} \\
      &   & L_{33}^{T} & L_{43}^{T} \\
      &   &   & L_{44}^{T} \\
\end{array}\right].
\]
This is followed by the second step, TILE\_TRTRI.
(See Algorithm~\ref{alg:InPlace} Step 2.)
It computes $T$, the inverse of the Cholesky factor such that
\[
\left[ \begin{array}[h]{cccc}
    T_{11} &   &   &   \\
    T_{21} & T_{22} &   &   \\
    T_{31} & T_{32} & T_{33} &   \\
    T_{41} & T_{42} & T_{43} & T_{44} \\
\end{array}\right]
=
\left[ \begin{array}[h]{cccc}
    L_{11} &   &   &   \\
    L_{21} & L_{22} &   &   \\
    L_{31} & L_{32} & L_{33} &   \\
    L_{41} & L_{42} & L_{43} & L_{44} \\
\end{array}\right]^{-1}.
\]
The third and last step is TILE\_LAUUM.  (See Algorithm~\ref{alg:InPlace} Step
3.) It multiplies $T$ with its transpose and provides $B$, the inverse of the
original matrix, $A$,
\[
\left[ \begin{array}[h]{cccc}
    T_{11}^{T} & T_{21}^{T} & T_{31}^{T} & T_{41}^{T} \\
      & T_{22}^{T} & T_{32}^{T} & T_{42}^{T} \\
      &   & T_{33}^{T} & T_{43}^{T} \\
      &   &   & T_{44}^{T} \\
\end{array}\right]
\left[ \begin{array}[h]{cccc}
    T_{11} &   &   &   \\
    T_{21} & T_{22} &   &   \\
    T_{31} & T_{32} & T_{33} &   \\
    T_{41} & T_{42} & T_{43} & T_{44} \\
\end{array}\right]
=
\left[ \begin{array}[h]{cccc}
    B_{11} & \cdot & \cdot & \cdot \\
    B_{21} & B_{22} & \cdot & \cdot \\
    B_{31} & B_{32} & B_{33} & \cdot \\
    B_{41} & B_{42} & B_{43} & B_{44} \\
\end{array}\right].
\]
This algorithm can be done in-place. In the sense that $L$ can be computed
in-place of $A$.  $T$ in-place of $L$ and $B$ in-place of $T$.

The individual tasks within each step is a BLAS or LAPACK sequential
functionality: POTRF, LAUUM, TRTRI, TRSM, TRMM, SYRK, or GEMM.
See~\cite{BientinesiGunterVanDeGeijn:08,Quintana:2009,vecpar10} for more
information on the tiled Cholesky inversion algorithm and in particular the
definition of variants 1, 2, and 3 for TILE\_TRTRI.

In this paper, we consider two weights for the tasks. Either we weight each
task equally, or  we weight each task according to the number of flops it
requires.  In Table~\ref{tab:taskweights}, we present the weights for each
tasks when we use the number of flops as metric.  We take one unit to be
$\frac{b^3}{3}$ flops. In this case, neglecting any lesser terms, the weight
of each task becomes a simple integer.

\begin{table}[htbp]
        \centering
        \begin{tabular}[ht]{ccc}
            \hline\hline
                  & \# flops  & flop-based weights (in $\frac{b^3}{3}$ flops)\\
            \hline
            POTRF & $\frac{1}{3}b^3$ & 1 \\
            LAUUM & $\frac{1}{3}b^3$ & 1 \\
            TRTRI & $\frac{1}{3}b^3+\frac{2}{3}b$ & 1 \\
            TRSM  & $b^3$            & 3 \\
            TRMM  & $b^3$ & 3 \\
            SYRK  & $b^3 + b$        & 3 \\
            GEMM  & $2b^3$           & 6 \\
            \hline
        \end{tabular}\par
    \caption{Task Weights}\label{tab:taskweights}
\end{table}

Although each of the three steps is distinct from each other, common to all
three is the total number of tasks and the total number of flops.  For each,
TILE\_POTRF, TILE\_TRTRI, and TILE\_LAUUM, the total number of tasks is
$\frac{1}{6}\left( t^3 + 3t^2 + 2t \right)$ and a total number of $t^3$ flops.

In this paper, we study different variants for our algorithms. In our analysis,
we consider a constant granularity (block size) for all algorithms. In this
framework, {\em we consider an algorithm better than another if it has a
shorter critical path.} We show the merit of this approach in our experimental
section.  We note that our analysis relies on appropriate choice for the
weights of each tasks.  Different choices of weights lead to different answers,
different critical path lengths, and, indeed, different critical paths.
Whether the point of view is to consider equal weights for each task or to
weight according to the number of flops for each task, pertinent information
about the performance of the algorithm can be extracted in either case.  Both
models can be found in the literature and are fairly standard.  Weighting each
task with their total number of flops is justified since our tasks perform
$\mathcal{O}(n^3)$ operations for $\mathcal{O}(n^2)$ data transfer.  Weighting
each task as one unit emphasizes the latency of starting a task and might model
some overhead associated with tasks (as data transfer).  Other weights are not
excluded but we only consider these two models in this manuscript.

The layout of the following sections will run somewhat counter intuitive with
respect to the steps of the algorithm and will instead follow the progression
of complexity of the steps.  We present the results for Step 1, the Cholesky
factorization (TILE\_POTRF), which is succeeded by Step 3, the matrix
multiplication (TILE\_LAUUM), followed by Step 2, the triangular inversion
(TILE\_TRTRI).  After which, the complete algorithm (CHOLINV) is taken into
account. 

\linesnumbered 
\begin{algorithm}
  \KwIn{$A$, Symmetric Positive Definite matrix in tile storage ($t\times t$ tiles).}
  \KwResult{$A^{-1}$, stored in-place in $A$.}
  \emph{Step~1: Tile Cholesky Factorization (compute L such that $A=LL^T$)}\;
  \For{$j=0$ \KwTo $t-1$}{
    \For{$k=0$ \KwTo $j-1$}{
      $A_{j,j} \leftarrow A_{j,j} - A_{j,k} \ast A_{j,k}^T$ (SYRK(j,k)) \;
    }
    $A_{j,j} \leftarrow CHOL(A_{j,j})$ (POTRF(j)) \;
    \For{$i=j+1$ \KwTo $t-1$}{
      \For{$k=0$ \KwTo $j-1$}{
        $A_{i,j} \leftarrow A_{i,j} - A_{i,k} \ast A_{j,k}^T$ (GEMM(i,j,k)) \;
      }
    }
    \For{$i=j+1$ \KwTo $t-1$}{
      $A_{i,j} \leftarrow A_{i,j} / A_{j,j}^T$ (TRSM(i,j)) \;
    }
  }
  \emph{Step~2: Tile Triangular Inversion of $L$ (compute $L^{-1}$)}\;
  \For{$j=t-1$ \KwTo $0$}{
    $A_{j,j} \leftarrow TRINV(A_{j,j})$ (TRTRI(j)) \;
    \For{$i=t-1$ \KwTo $j+1$}{
      $A_{i,j} \leftarrow A_{i,i} \ast A_{i,j}$ (TRMM(i,j)) \;
      \For{$k=j+1$ \KwTo $i-1$}{
        $A_{i,j} \leftarrow A_{i,j} + A_{i,k} \ast A_{k,j} $ (GEMM(i,j,k)) \;
      }
      $A_{i,j} \leftarrow - A_{i,j} \ast A_{i,i}$ (TRMM(i,j)) \;
    }
  }
  \emph{Step~3: Tile Product of Lower Triangular Matrices (compute $A^{-1}={L^{-1}}^TL^{-1}$)}\;
  \For{$i=0$ \KwTo $t-1$}{
    \For{$j=0$ \KwTo $i-1$}{
      $A_{i,j} \leftarrow A_{i,i}^T \ast A_{i,j}$ (TRMM(i,j)) \;
    }
    $A_{i,i} \leftarrow A_{i,i}^T \ast A_{i,i}$ (LAUUM(i)) \;
    \For{$j=0$ \KwTo $i-1$}{
      \For{$k=i+1$ \KwTo $t-1$}{
        $A_{i,j} \leftarrow A_{i,j} + A_{k,i}^T \ast A_{k,j}$ (GEMM(i,j,k)) \;
      }
    }
    \For{$k=i+1$ \KwTo $t-1$}{
      $A_{i,i} \leftarrow A_{i,i} + A_{k,i}^T \ast A_{k,i}$ (SYRK(i,k)) \;
    }
  }
  \caption{Tile In-place Cholesky Inversion (lower format). Matrix
    $A$ is the on-going updated matrix (in-place algorithm).}
  \label{alg:InPlace}
\end{algorithm}

\section{Analysis of Cholesky Factorization - TILE\_POTRF}
\begin{figure}
  \begin{tabular}{@{}p{.45\linewidth}p{.10\linewidth}p{.45\linewidth}@{}}
    \centering
    \subfigure[Tasks perspective (equal weights)]{
      \label{fig:POTRFtasks}
      \includegraphics[width=.45\textwidth]{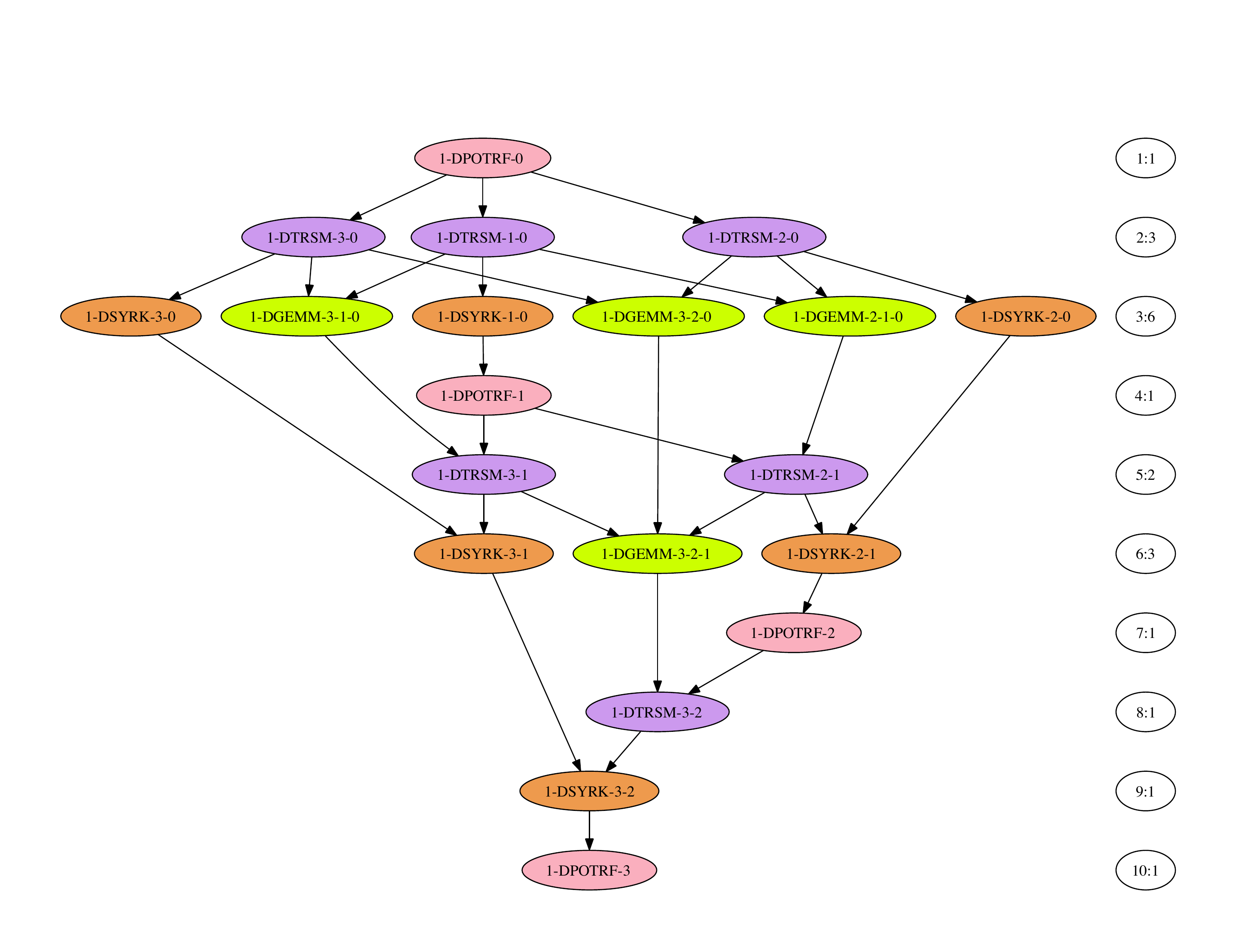}
    }%
    &
    &
    \centering
    \subfigure[Flops perspective (unequal weights)]{
      \label{fig:POTRFflops}
      \includegraphics[width=.45\textwidth]{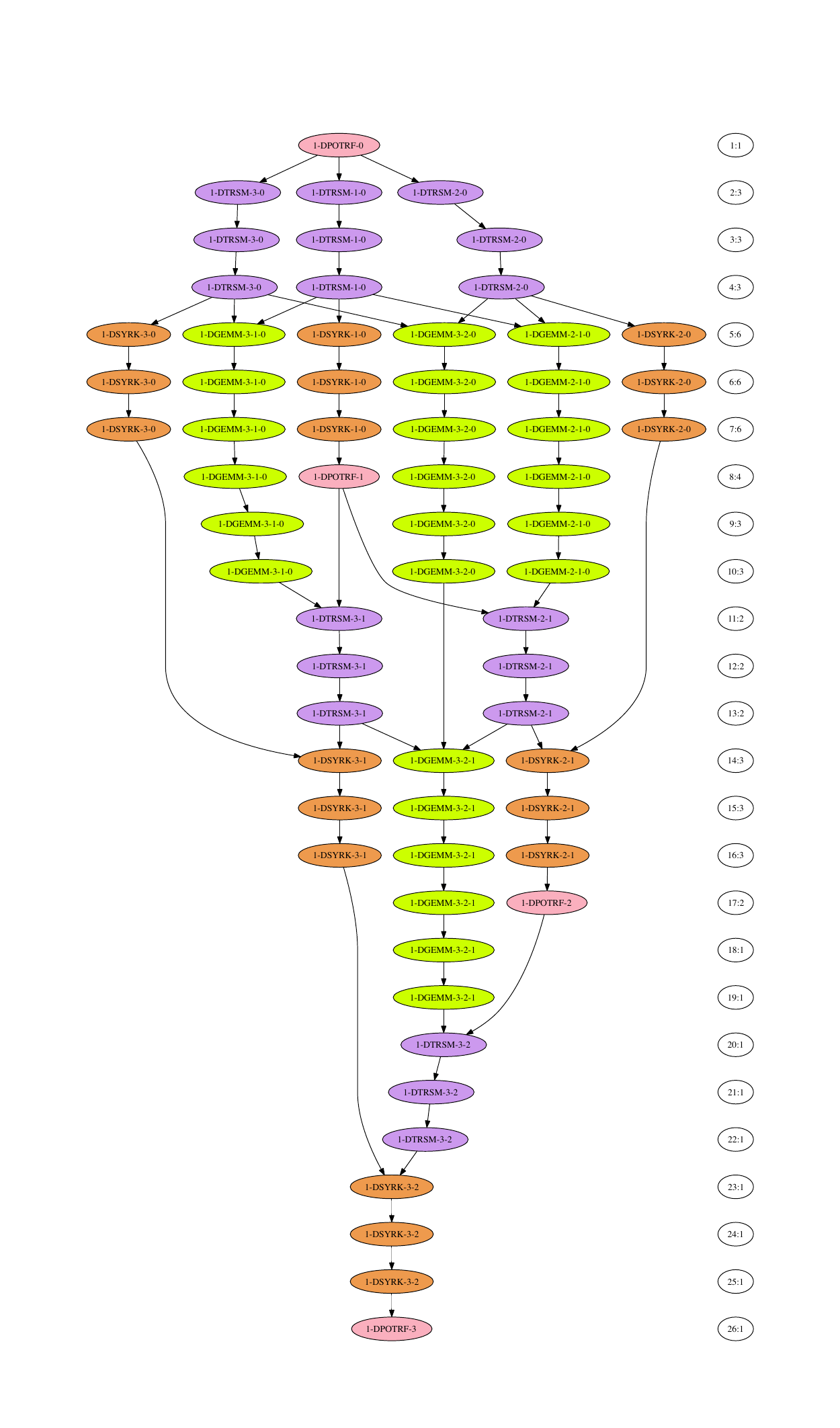}
    }%
  \end{tabular}
  \caption{TILE\_POTRF DAGs for tasks and flops ($t=4$).}
  \label{fig:POTRFdags}
\end{figure}

In the first step, the Cholesky factorization of an $n \times n$ real symmetric
positive definite matrix $A$ can be of the form $LL^{T}$, where $L$ is a lower
triangular matrix having positive elements on the diagonal.  Albeit that there
are three variants of the Cholesky factorization (bordered, right-looking,
left-looking), the DAGs produced are all identical and are represented, for
$t=4$, in Figure~\ref{fig:POTRFtasks}.

In view of the tasks weighted equally, the critical path follows POTRF, TRSM,
and SYRK for each $t-1$ with another POTRF at the final step (refer
to Figure~\ref{fig:POTRFtasks}.  Hence the length of the critical path is a linear
function:

\[ (1 + 1 + 1)(t - 1) + 1 = 3t -2 \]

Analogously, the flops follow POTRF, then TRSM and GEMM for each $t-2$ with
another TRSM, SYRK and POTRF at the final step resulting in a linear function:
\[ 1 + (3 + 6)(t - 2) + 3 + 3 + 1 = 9t -10 \]
(refer to Figure~\ref{fig:POTRFflops}).  Table~\ref{tab:cpPOTRF} describes each
of these equations as a function of $t$.

\begin{table}[hbt]
        \centering
        \begin{tabular}{ccc}
            \hline\hline
                  & Tasks  & Flops\\
            \hline
            TILE\_POTRF & $3t-2$ & $9t-10$ \\
            \hline
        \end{tabular}\par
    \caption{TILE\_POTRF critical path length}\label{tab:cpPOTRF}
\end{table}

TILE\_POTRF is an example where the critical path is changed whether we consider
flops-based weights or tasks-based weights.

\section{Analysis of Triangular matrix multiplication - TILE\_LAUUM}
\begin{figure}
    \begin{tabular}{@{}p{.45\linewidth}p{.10\linewidth}p{.45\linewidth}@{}}
        {\centering
        \subfigure[Tasks perspective (equal weights, in-place)]{
        \label{fig:LAUUMtasks}
        \includegraphics[width=.45\textwidth]{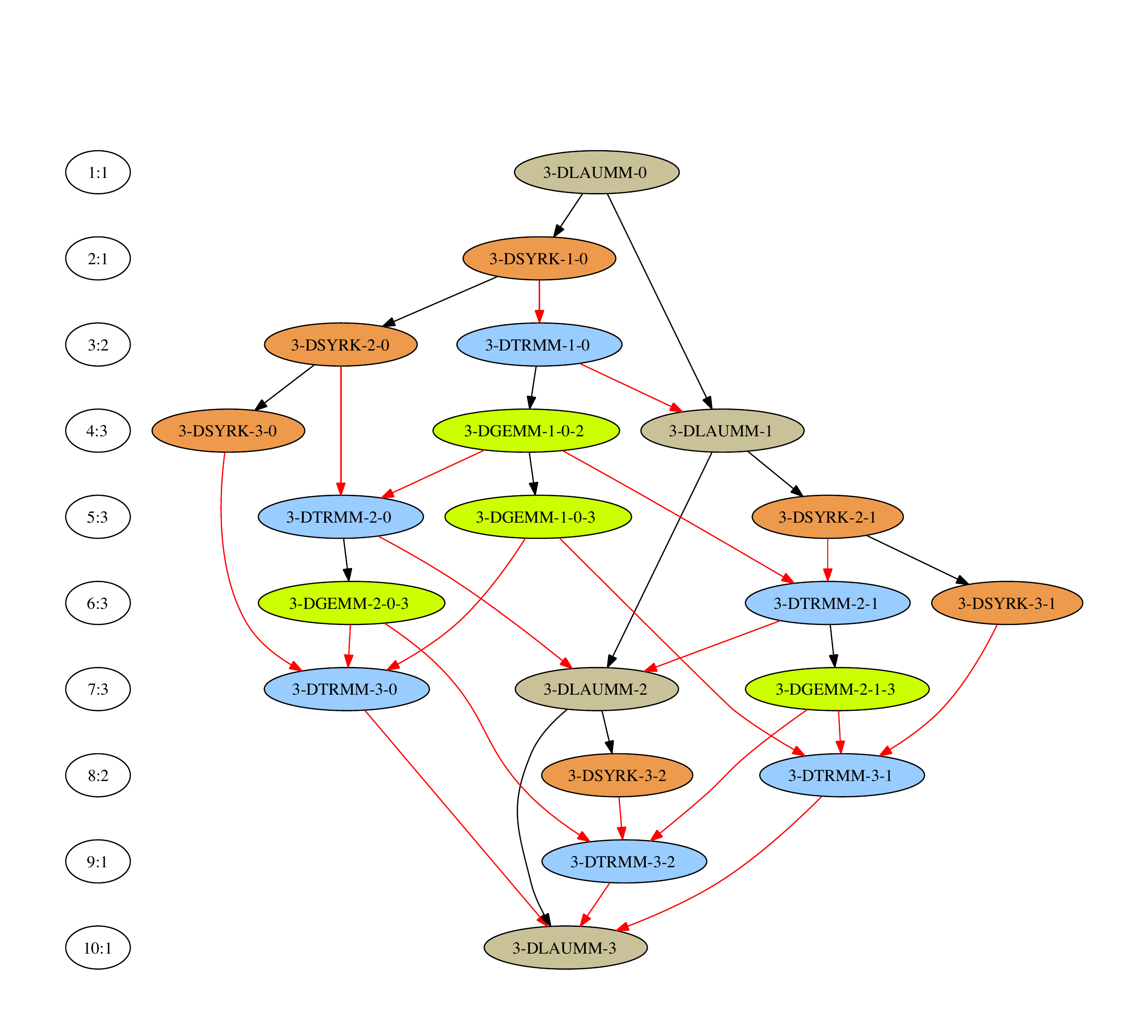}
        }}%
        & &
        {\centering
        \subfigure[Flops perspective (unequal weights, in-place)]{
        \label{fig:LAUUMflops}
        \includegraphics[width=.45\textwidth]{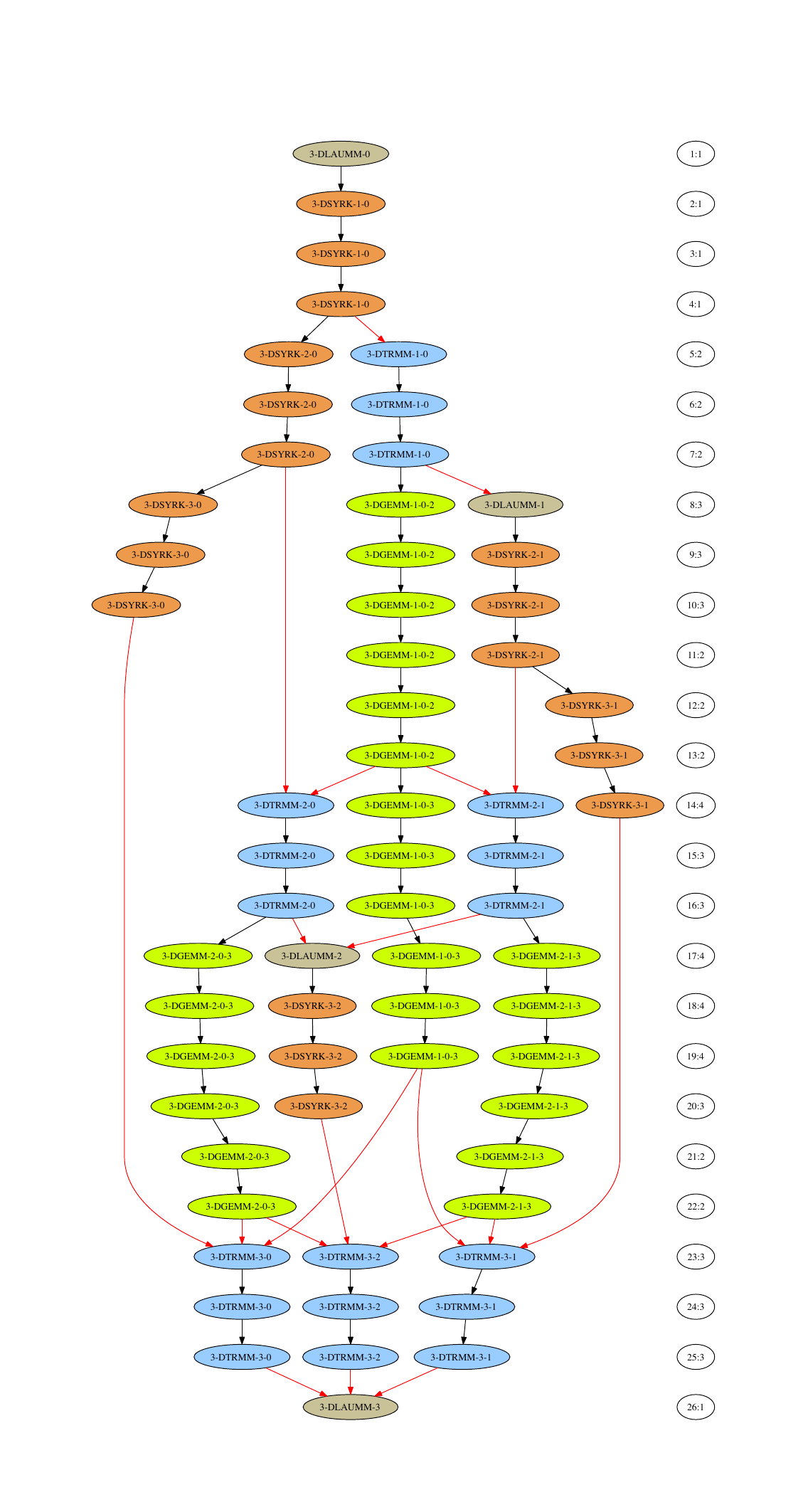}
        }} \\
        {\centering
        \subfigure[Tasks perspective (equal weights, out-of-place)]{
        \label{fig:LAUUMtasksout}
        \includegraphics[width=.45\textwidth]{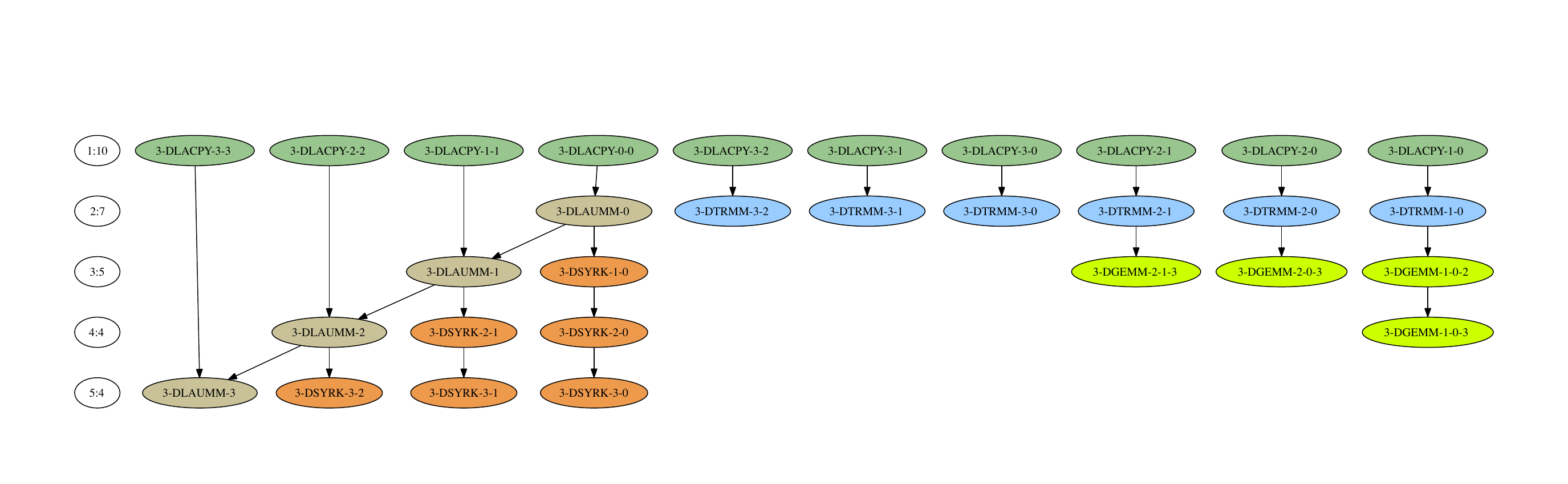}
        }}%
        & & 
        {\centering
        \subfigure[Flops perspective (unequal weights, out-of-place)]{
        \label{fig:LAUUMflopsout}
        \includegraphics[width=.45\textwidth]{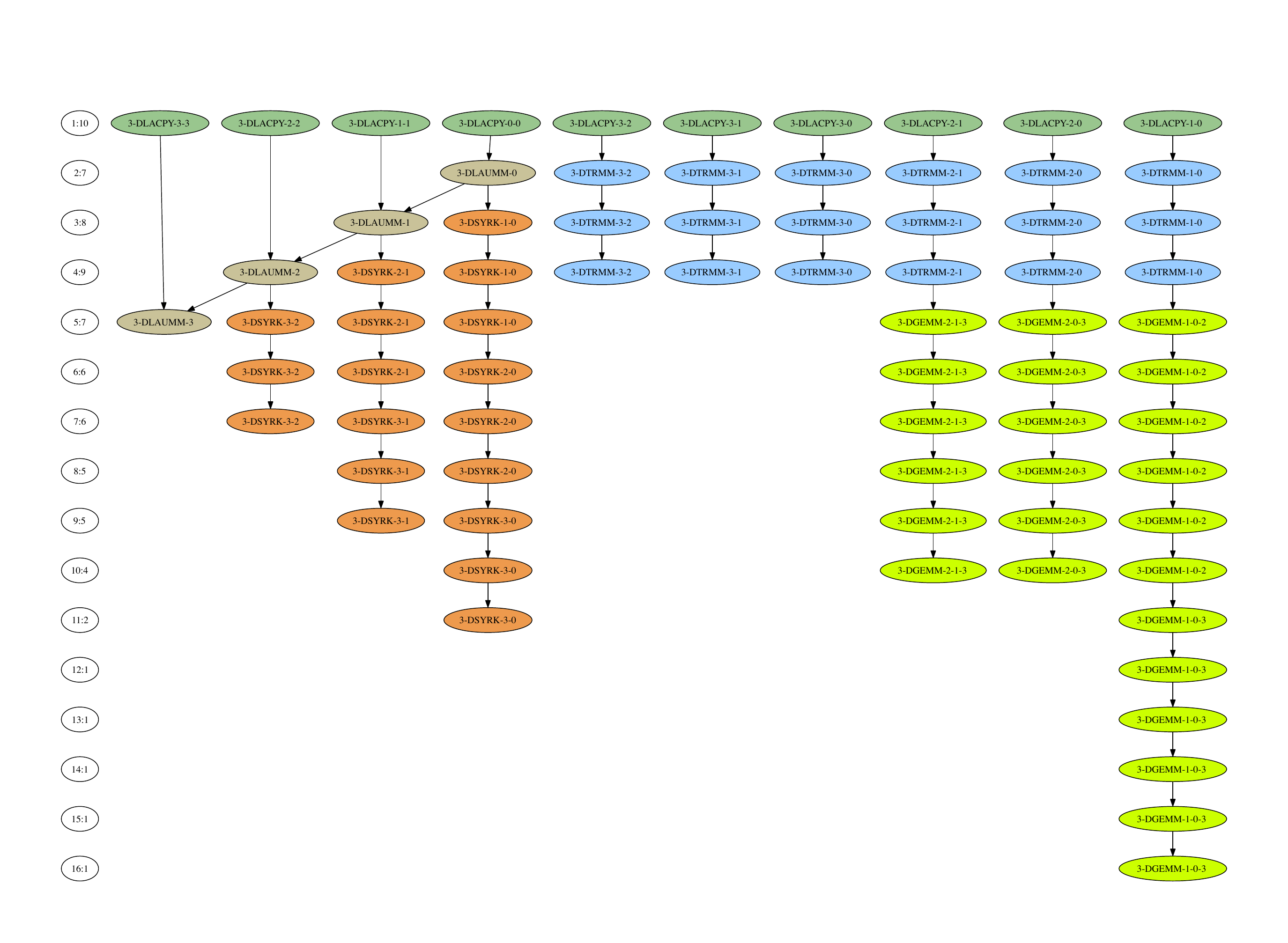}
        }}%
    \end{tabular}
    \caption{TILE\_LAUUM DAGs for  $t=4$, in-place and out-of-place.}
    \label{fig:LAUUMdagsout}
\end{figure}

As with the first step, the third step can have multiple variants dependent
upon the order the result computed, either column or row wise, but the
resulting DAGs are all identical (Figure~\ref{fig:LAUUMtasks}); it is simply a
multiplication of two triangular matrices.  However, since the result is stored
in-place, there are many dependencies arising from a write-after-read (WAR)
operation.  In order to break this dependence, a buffer must be used to allow
multiple operations to read a particular tile while another operation over
writes it; we call the variants without buffer `in-place' and those using
buffers `out-of-place'.  In so doing, the DAG changes dramatically as shown in
Figure~\ref{fig:LAUUMtasks} and Figure~\ref{fig:LAUUMtasksout}. The cost of
using the buffer is considered as one unit (whether we are flops-based or
tasks-based) and is incorporated into the DAG. In either case, the lengths of
the critical paths for both tasks-based and flops-based is linear in $t$
(Table~\ref{tab:cpLAUUM}).  

For the in-place variants, the critical path for the unweighted tasks follows
LAUUM, SYRK, TRSM for $t-1$ with a final LAUMM at the end such that the length
in terms of $t$ becomes:
\[ (1 + 1 + 1)(t - 1) + 1 = 3t - 2 \]
and for the weighted tasks the critical path follows LAUUM, SYRK with TRSM and
GEMM for $t-2$, and TRSM and LAUUM bringing up the end:
\[ 1 + 3 + (3 + 6)(t - 2) + 3 + 1 = 9t -10. \]
For the out-of-place condition, we have for the unweighted tasks a critical
path of LAUUM followed by $t-1$ SYRKs and the cost of using the buffer:
\[ 1 + (t - 1) + 1 = t + 1. \]
Observe that for the weighted tasks, the out-of-place critical path follows
TRSM and $t-2$ GEMMs and the cost of using the buffer for values of $t \geq 3$
(for $t=2$, we would have $3t-1$):
\[ 3 + 6 (t - 2) + 1 = 6t - 8. \]
All of these are summarized in Table~\ref{tab:cpLAUUM}.

\begin{table}
        \centering
        \begin{tabular}{ccc}
            \hline\hline
                  & Tasks  & Flops\\
            \hline
            TILE\_LAUUM (in-place) & $3t-2$ & $9t-10$ \\
            TILE\_LAUUM (out-of-place) & $t+1$ & $6t-8$ \\
            \hline
        \end{tabular}\par
    \caption{TILE\_LAUUM critical path length ($t \geq 3$)}\label{tab:cpLAUUM}
\end{table}

\section{Analysis of Triangular matrix inversion - TILE\_TRTRI}\label{sec:TRTRI}
Of the three steps, the triangular inversion provides the most interest.  The
six variants that we have studied can be grouped into two groups of three by
consideration of the mathematical approach, either by using the left inverse
$T^{-1}T = I$ or the right inverse $TT^{-1} = I$;  variants 1 through 3 use the
left inverse and 4 through 6 use the right inverse.  The left inverse moves
through the matrix from the upper left corner to the lower right and vice versa
for the right inverse.  Thus, when speaking of the DAGs and critical paths, we
will focus on one group since the other group is similar.  As with the
triangular matrix multiplication, we consider both in-place variants  
and out-of-place variants, which break the some of the WAR dependencies.

\begin{figure}
    \begin{tabular}{@{}p{.475\linewidth}p{.05\linewidth}p{.475\linewidth}@{}}
        \multicolumn{3}{c}{
    \centering
    \subfigure[TILE\_TRTRI v1]{
      \label{fig:TRTRIdag_v1}
      \includegraphics[height=4cm, width=.225\textwidth]{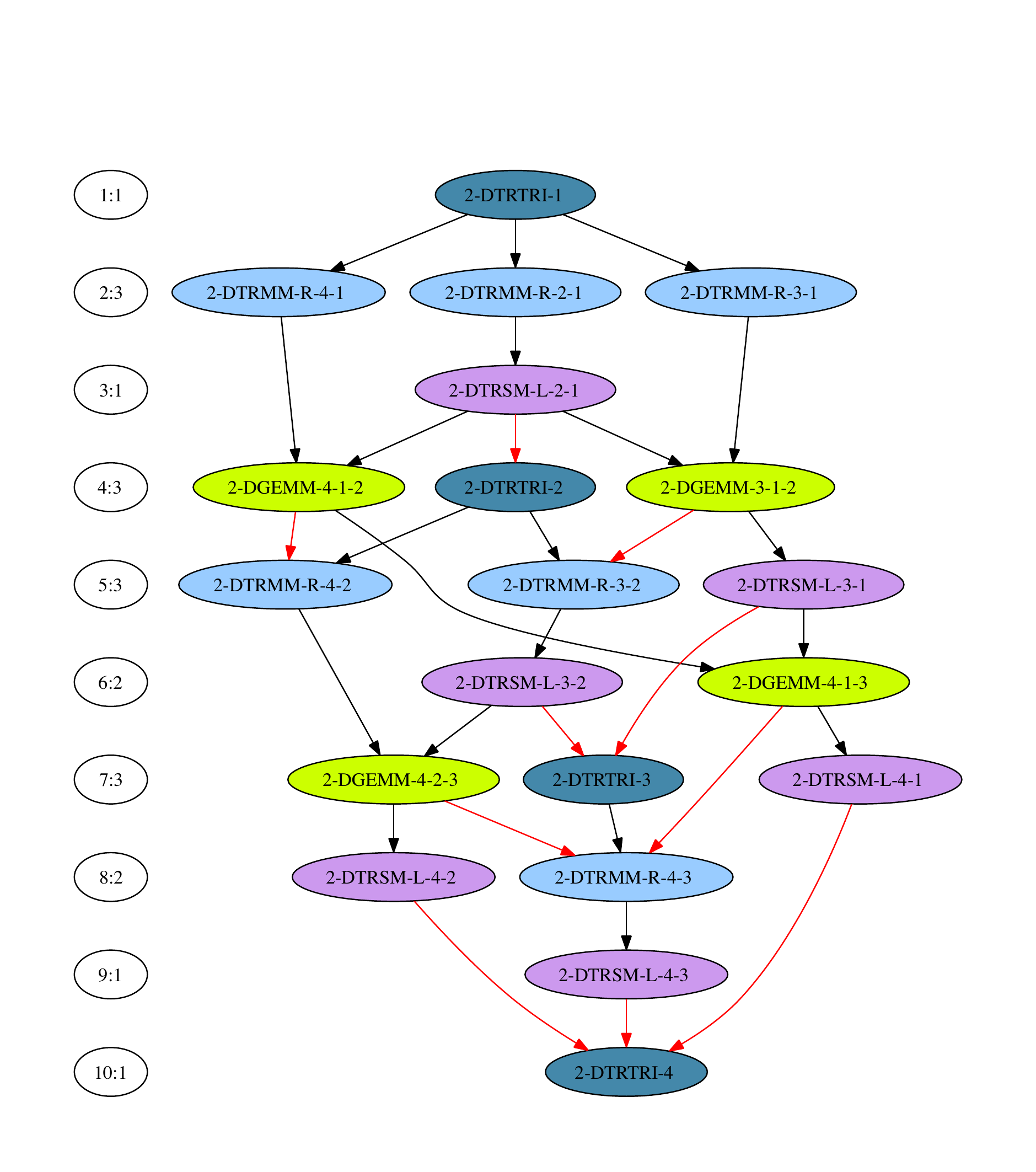}
      }}\\
    \centering
    \subfigure[TILE\_TRTRI v2]{
      \label{fig:TRTRIdag_v2}
      \includegraphics[height=3cm, width=.35\textwidth]{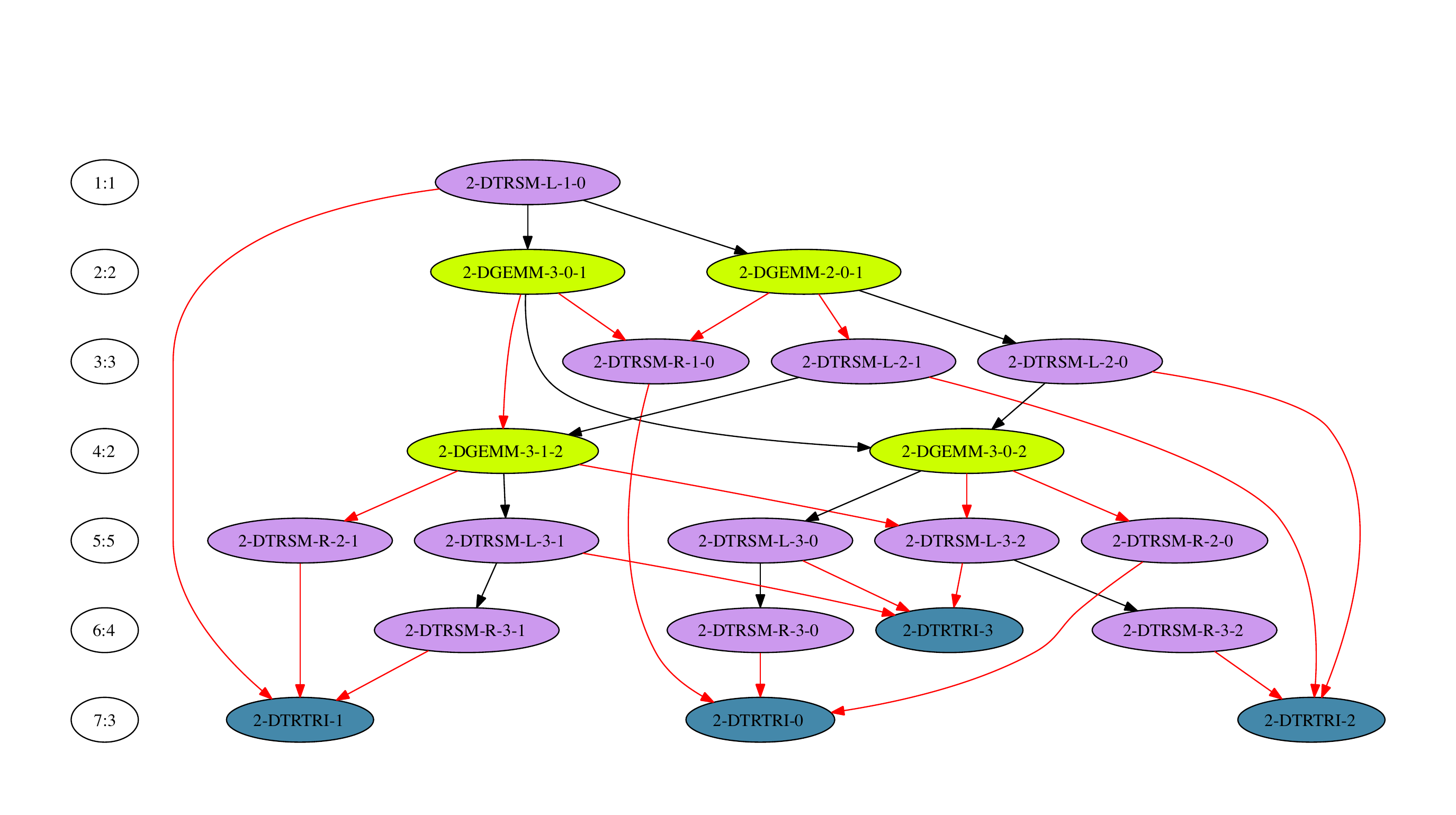}
    }%
    & &
    \centering
    \subfigure[TILE\_TRTRI v3]{
      \label{fig:TRTRIdag_v3}
      \includegraphics[height=2.5cm, width=.42\textwidth]{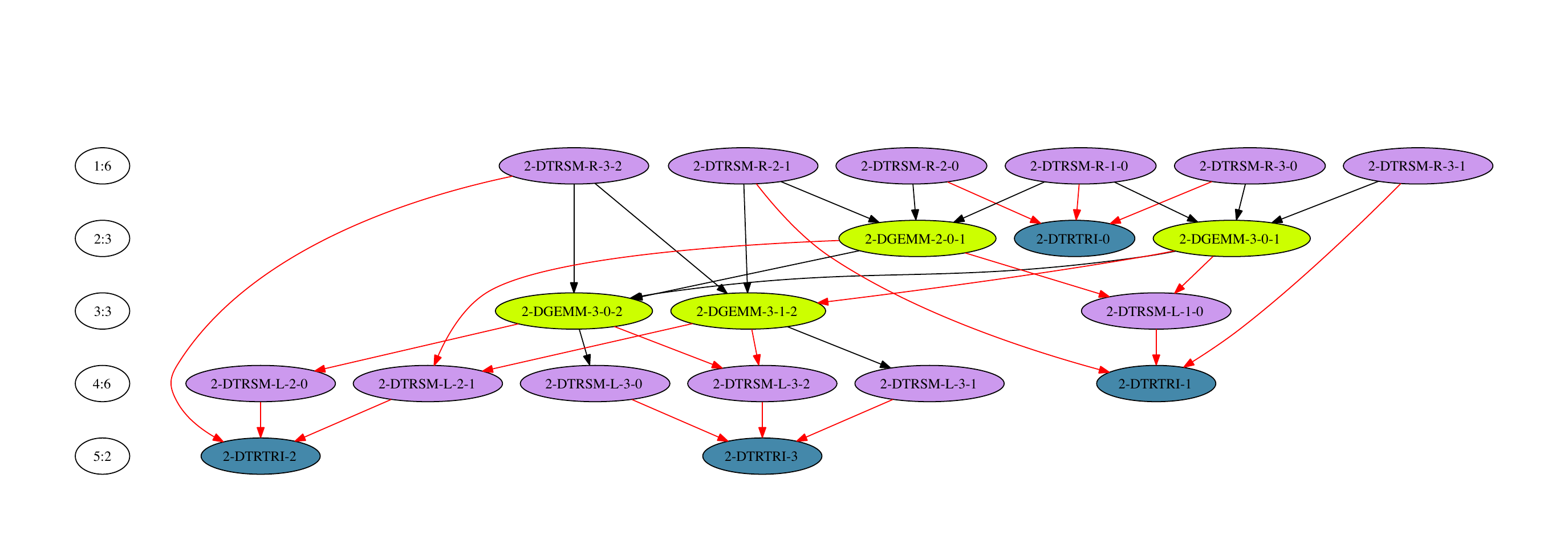}
    }%
  \end{tabular}
  \caption{DAGs for three variants of TILE\_TRTRI ($t=4$) in-place.}
  \label{fig:TRTRIdags}
\end{figure}

\begin{figure}
    \begin{tabular}{@{}p{.475\linewidth}p{.05\linewidth}p{.475\linewidth}@{}}
        \multicolumn{3}{c}{
        \subfigure[TILE\_TRTRI v1]{
        \label{fig:TRTRIdag_v1_op}
        \includegraphics[width=.475\textwidth]{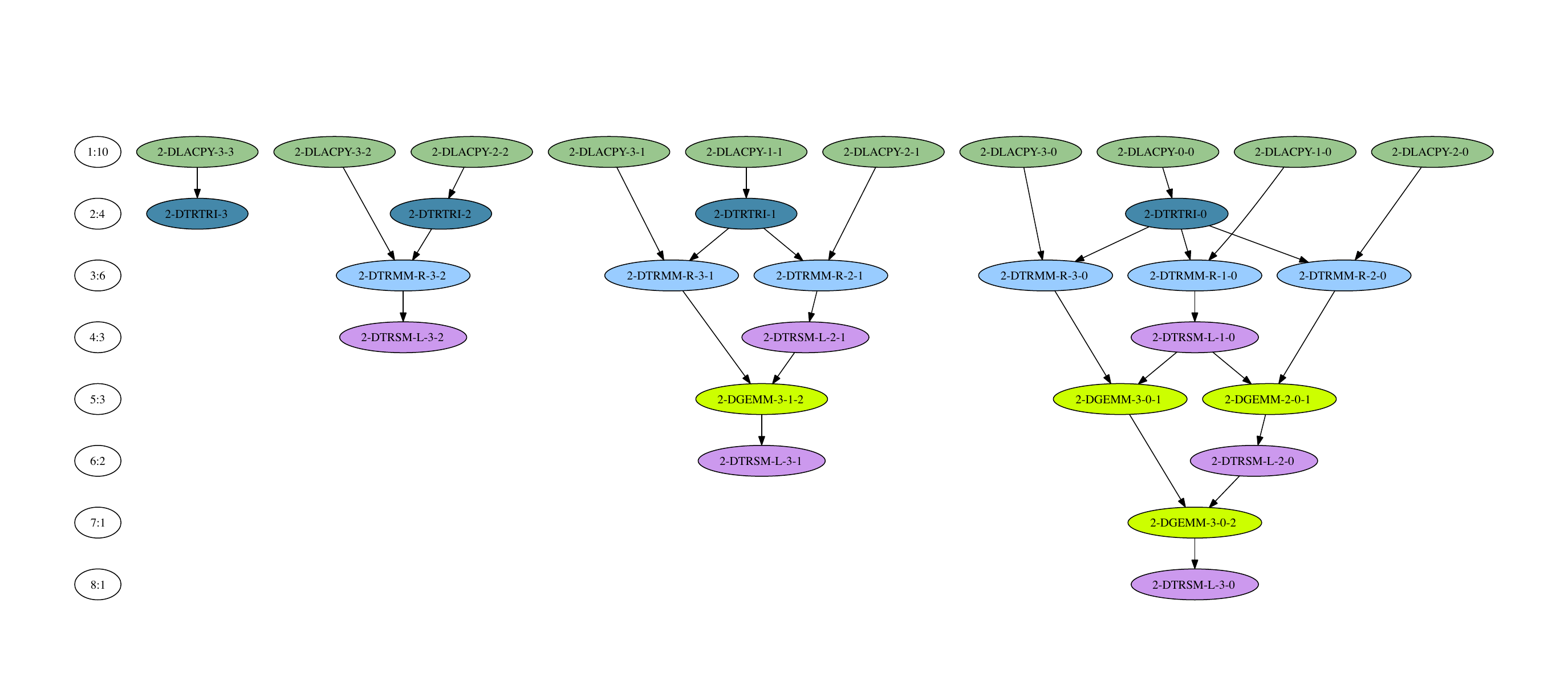}
      }}\\
    \centering
    \subfigure[TILE\_TRTRI v2]{
      \label{fig:TRTRIdag_v2_op}
      \includegraphics[width=.475\textwidth]{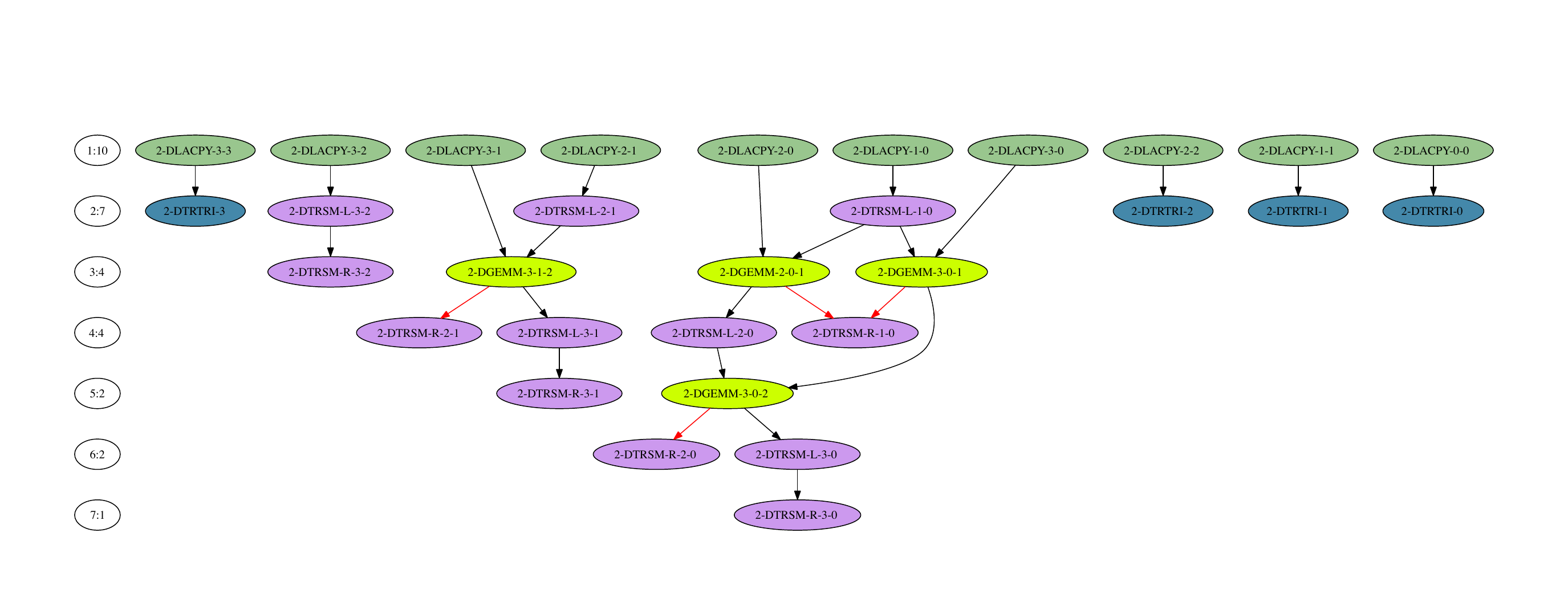}
    }%
    & &
    \centering
    \subfigure[TILE\_TRTRI v3]{
      \label{fig:TRTRIdag_v3_op}
      \includegraphics[width=.475\textwidth]{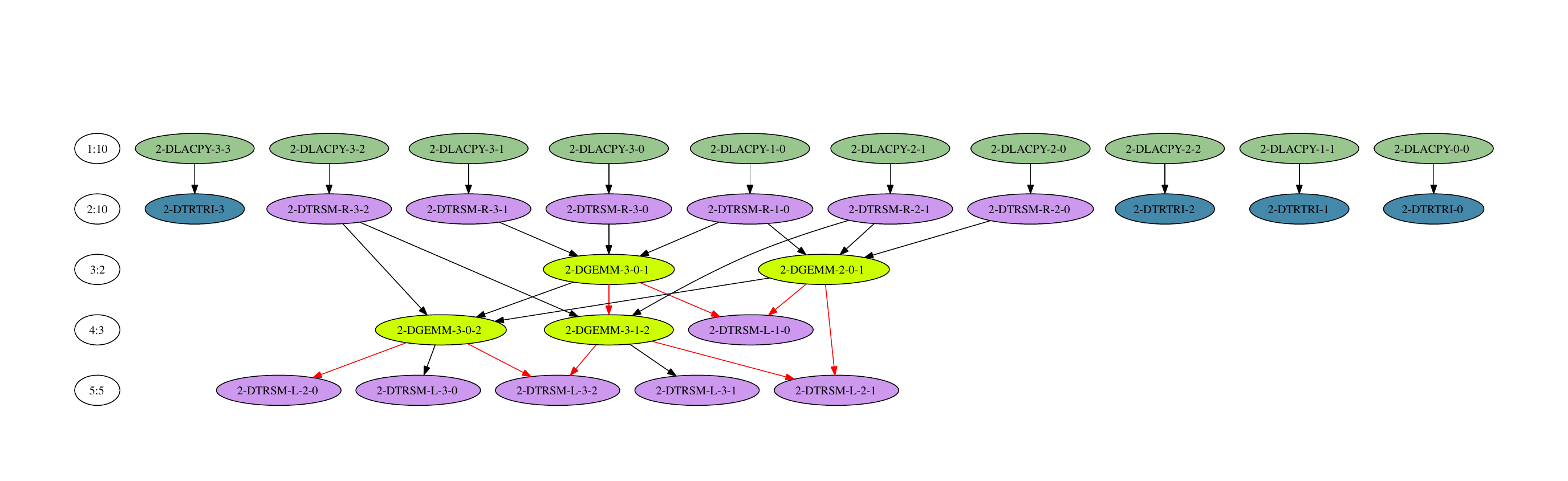}
    }%
  \end{tabular}
  \caption{DAGs for three variants of TILE\_TRTRI ($t=4$) out-of-place.}
  \label{fig:TRTRIdagsout}
\end{figure}

Unlike in the previous sections, the DAGs for the three variants, for both
in-place and out-of-place, are not identical as can be seen in tasks viewpoint
in Figures~\ref{fig:TRTRIdags} and~\ref{fig:TRTRIdagsout}.

As before, the lengths of the critical paths for the tasks and the flops are
linear function of $t$ and are provided in Table~\ref{tab:cpTRTRI}.  Note that
although the in-place and out-of-place DAGs are different for a single variant,
only variant 1 reaps any benefit from the use of the buffers.  For the others,
the cost of providing the buffer, which is considered to be one unit, negates
any advantage it may provide.

In the unweighted case, we look at the lengths of the critical paths for the
tasks.  For variant 1, the critical path traverses $t-1$  TRTRI, TRMM and TRSMs
ending with TRTRI.  Thus
\[ (1 + 1 + 1) (t - 1)  + 1 = 3t - 2.\]
For variant 2, the critical path traverses TRTRI followed by $t-2$ GEMM
and TRSMs and ends with a TRSM and a TRTRI.  Thus
\[ 1 + (1 + 1)(t - 2) + 1 + 1 = 2t - 1.\]
For variant 3, the critical path traverses TRTRI followed by $t-2$ GEMMs and
ends with a TRSM and a TRTRI.  Thus
\[ 1 + (t - 2) + 1 + 1 = t + 1.\]

Similarly, in the weighted case we consider the critical path of each variant.
For variant 1, the critical path traverses TRTRI followed by $t-2$ TRMM, TRSM
and GEMMs and ends with a TRMM, TRSM and a TRTRI.  Thus
\[ 1 + (3 + 3 + 6)(t - 2) + 3 + 3 + 1 = 12t - 16.\]
For variant 2, the critical path traverses TRTRI, followed by $t-2$ GEMM and
TRSMs ending with a TRSM and TRTRI.  Thus
\[ 3 + (6 + 3) (t - 2)  + 3 +  1 = 9t - 11.\]
For variant 3, the critical path traverses TRTRI followed by $t-2$ GEMMs and
ends with a TRSM and a TRTRI.  Thus
\[ 3 + 6(t - 2) + 3 + 1 = 6t - 5.\]
All of the above results are summarized in Table~\ref{tab:cpTRTRI}.

\begin{table}
        \centering
        \begin{tabular}[ht]{cccc}
            \hline\hline
                  & Variant & Tasks  & Flops\\
            \hline
            \multirow{3}{*}{TRTRI (in-place)} & 1,4 & $3t-2$ & $12t-16$ \\
                                              & 2,5 & $2t-1$ & $9t-11$ \\
                                              & 3,6 & $t+1$  & $6t-5$ \\
            \hline
            \multirow{3}{*}{TRTRI (out-of-place)} & 1,4 & $2t$ & $9t-10$ \\
                                              & 2,5 & $2t-1$ & $9t-11$ \\
                                              & 3,6 & $t+1$  & $6t-5$ \\
            \hline
        \end{tabular}\par
    \caption{TRTRI critical path length}\label{tab:cpTRTRI}
\end{table}

\section{Analysis of Cholesky inversion - CHOLINV}
By combining the above three steps, we are able to compute the inverse of an
SPD matrix.  One approach is to perform the steps in sequential order such that
each step is not started until the previous step has been completed fully.
However, more parallelism can be obtained by interleaving the above three steps
while still adhering to any dependencies that exist among tasks either within
the step or between the steps and being cognizant of which variants are chosen
to maximize the interleaving. 

If one naively combines any variant of the TILE\_POTRF with variants 4 through
6 of TILE\_TRTRI, due to the fact that TILE\_POTRF moves from upper left to
lower right and these variants of TILE\_TRTRI move from lower right to upper
left, a sequential algorithm in terms of the steps is obtained.  Furthermore,
combining this with any of the variants of TILE\_LAUUM would result in a
completely sequential algorithm for the Cholesky inversion. We will see that
indeed variants 1 through 3 for the TILE\_TRTRI provide better theoretical and
experimental results as we would expect.

For each of the interleaved variants, we continue to observe the linear
behavior of the critical path in terms of tasks and flops as seen in
Table~\ref{tab:cpCHOLINV}.  Of particular interest is that the combination with
variant 1 of TILE\_TRTRI leads to a critical path length, in terms of tasks, of four
more tasks for the entire inversion ($3t+2$) as compared to just the Cholesky
factorization ($3t-2$), independent of the number of tiles. This is quite a feat. 

\begin{table}
    \centering
    \begin{tabular}[htb]{cccc}
        \hline\hline
                                                & Variant & Tasks  & Flops    \\
        \hline
        \multirow{6}{*}{CHOLINV (in-place)}     & x1x     & $3t+2$ & $12t+ 2$ \\
                                                & x2x     & $6t-1$ & $18t-11$ \\
                                                & x3x     & $3t+6$ & $ 9t+23$ \\
                                                & x4x     & $9t-6$ & $30t-36$ \\
                                                & x5x     & $8t-7$ & $27t-34$ \\
                                                & x6x     & $7t-3$ & $24t-25$ \\
        \hline
        \multirow{6}{*}{CHOLINV (out-of-place)} & x1x     & $3t+2$ & $ 9t+ 1$ \\ 
                                                & x2x     & $3t+2$ & $ 9t+ 7$ \\
                                                & x3x     & $3t+3$ & $ 9t+11$ \\
                                                & x4x     & $5t  $ & $18t-14$ \\
                                                & x5x     & $5t-3$ & $18t-19$ \\
                                                & x6x     & $5t-2$ & $21t-24$ \\
        \hline
    \end{tabular}
    \caption{CHOLINV critical path length} \label{tab:cpCHOLINV}
\end{table}

\begin{figure}
    \centering
    \includegraphics[width=.75\textwidth]{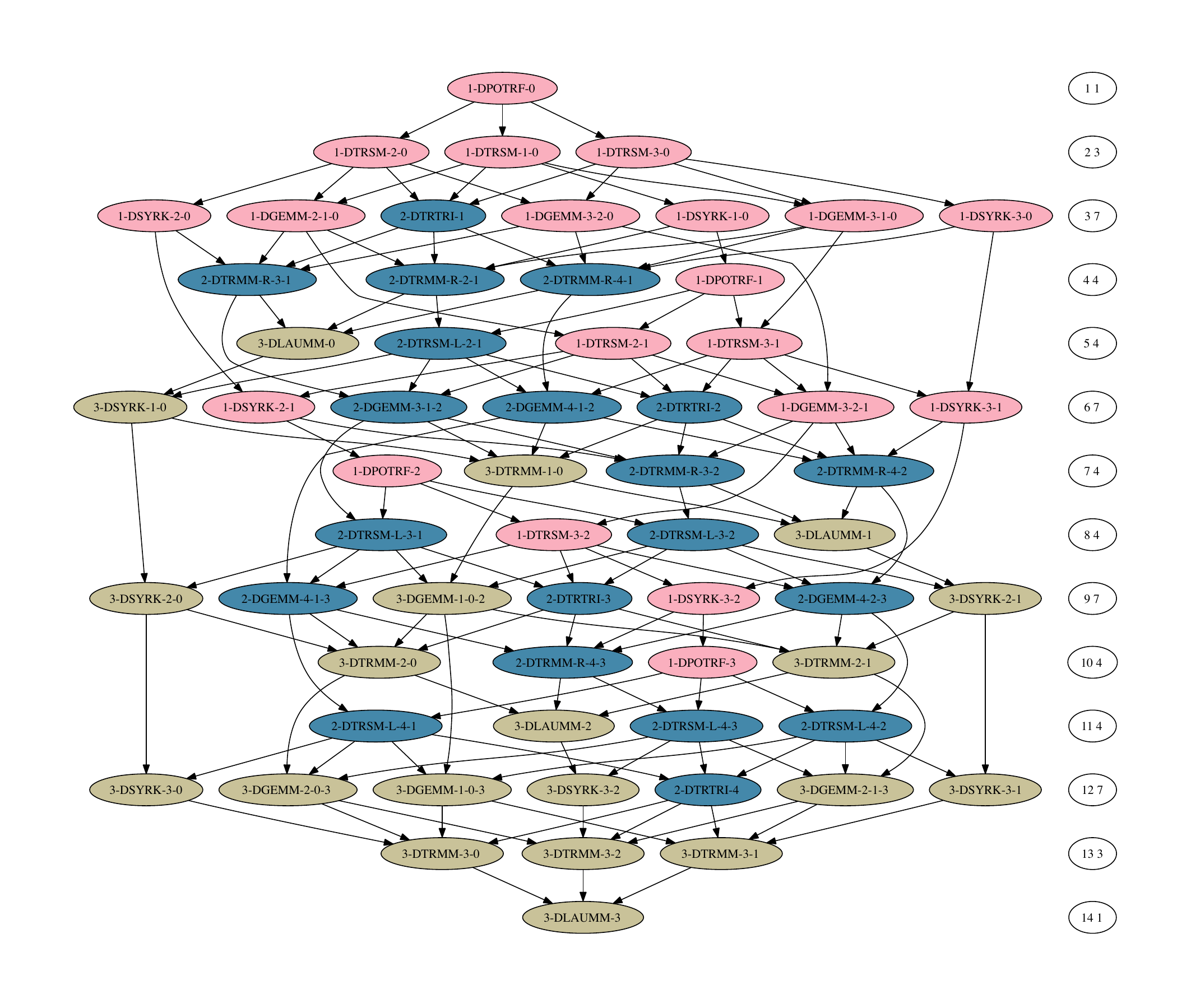}
    \caption{DAG for CHOLINV using variant 1 of TILE\_TRTRI ($t=4$) in-place.}
    \label{fig:CHOLINV311dag}
\end{figure}

Depicted in Figure~\ref{fig:CHOLINV311dag} is the Cholesky inversion, for four
tiles, using variant 1 of TILE\_TRTRI.  Each step is identified by a different
color to clearly see how the three steps are interleaving with each other.
This combination of variants allows portions of TILE\_TRTRI to start very early
on within Step 1 as well as portions of TILE\_LAUUM. One can observe the large
amount of parallelism obtained by the interleaving of the three steps. 
We see that the whole Cholesky inversion as three times more tasks as the Cholesky factorization
but finishes only 4 steps after.

With variant 3 of TILE\_TRTRI, the flops-based critical path for the Cholesky
inversion is the shortest. It is $9t+23$ which is only $33 \frac{b^3}{3}$ flops more
than the Cholesky factorization ($9t-10$). The difference between factorization and inversion is a constant number
independent of the number of tiles. Once more, this is quite a feat. 
These observations are in complete contrast with an analysis based on the total number of flops. 
The total number of flops for Cholesky inversion is three times more than the Cholesky factorization.
The tasks-based critical path lengths (using the appropriate variants) are about the same.

\section{Application of critical path analysis. Upper bound on performances.}

Having a closed form equation for the length of the critical path and knowing
the total number of flops for the entire algorithm, we can provide a lower
bound on the time to solution with the following reasoning: the total
execution time is at least the number of the flops on the critical path times
the flop rate ($\gamma$, in sec per flops), and it is at least the total number
of flops divided by the number of processors times the flop rate. This lower
bound on the execution time gives us an upper bound, $U(p)$, on the maximum
performance with $p$ cores. We obtain
\[ U(p) = \frac{1}{\gamma}\cdot \frac{\mbox{total number of flops}}{\max( \mbox{flops-based critical path length}, \frac{\mbox{total number of flops}}{p}  )} .\]
So that
\[ U(p) = \frac{1}{\gamma}\cdot \min( p, \frac{\mbox{total number of flops}}{\mbox{flops-based critical path length}}) .\]

\section{Experimental validation}

Our experiments were performed on an AMD Istanbul machine. This is a 48-core
machine which is composed of eight hexa-core Opteron 8439 SE (codename
Istanbul) processors running at 2.8 GHz. Each core has a theoretical peak of
11.2 Gflop/s with a peak of 537.6 Gflop/s for the whole machine. The Istanbul
micro-architecture is a NUMA architecture. Each socket has 6 MB of level-3
cache. Each processor has a 512 KB level-2 cache and a 128 KB level-1 cache.
After having benchmarked the AMD ACML and Intel MKL BLAS libraries, we selected
MKL (10.2) since it appeared to be slightly faster in our experimental context.
Linux 2.6.32 and Intel Compilers 11.1 were also used.

The sequential performance is taken as: 6.43 Gflop/s. This is obtained by
looking at a run on five or more cores and looking at the best achieved
performance of the kernels in this configuration.  Each core is able to perform
11.2 Gflop/s, so we estimate that our kernels are running at 57\% of the peak.

\begin{figure}
    \begin{tabular}{@{}p{.495\linewidth}p{.01\linewidth}p{.495\linewidth}@{}}
    \centering
    \subfigure[TILE\_TRTRI inplace]{
      \label{fig:TRTRI_threads_inplace}
      \includegraphics[width=.495\textwidth]{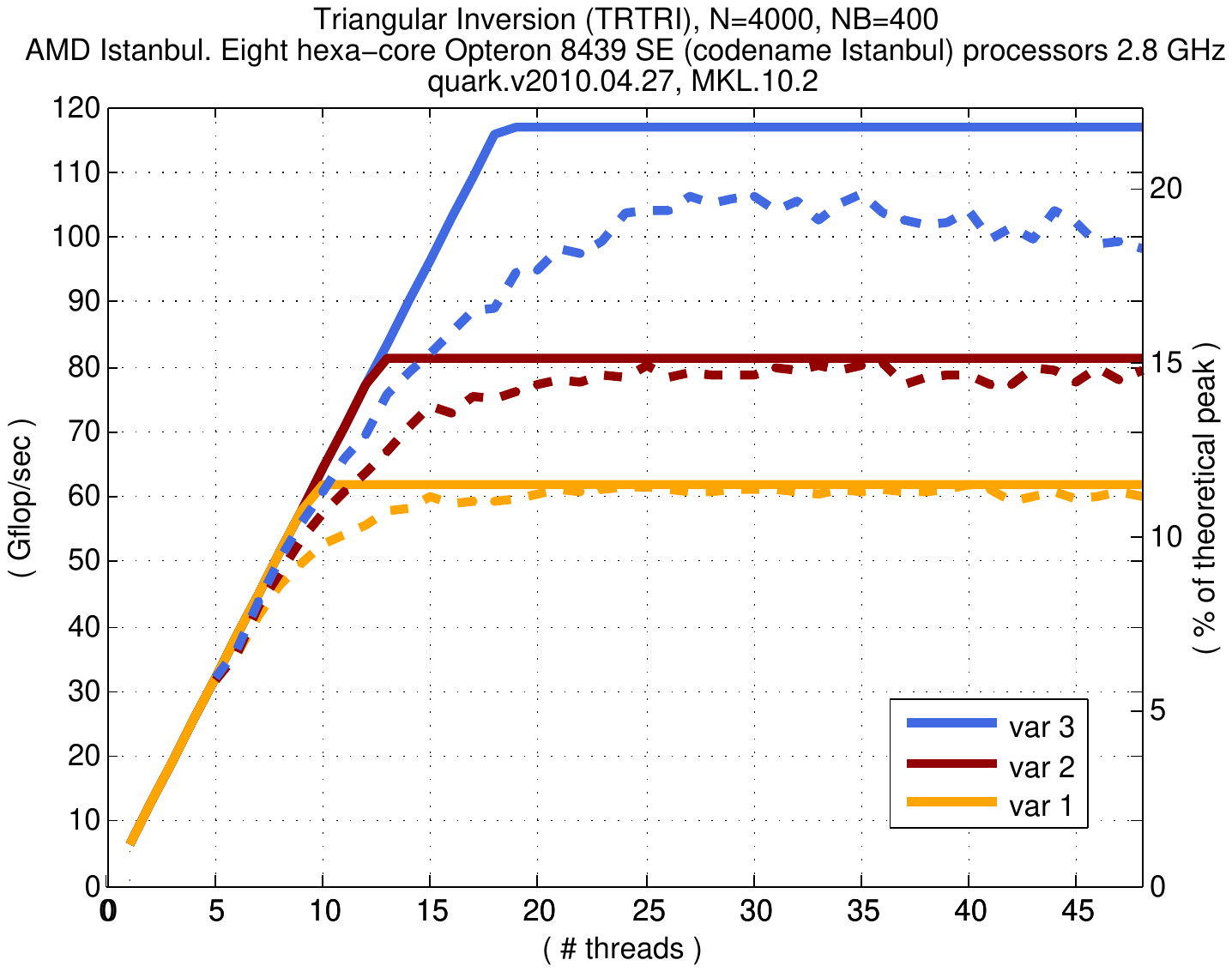}
      }
      & &
    \centering
    \subfigure[TILE\_TRTRI outofplace]{
      \label{fig:TRTRI_threads_outofplace}
      \includegraphics[width=.495\textwidth]{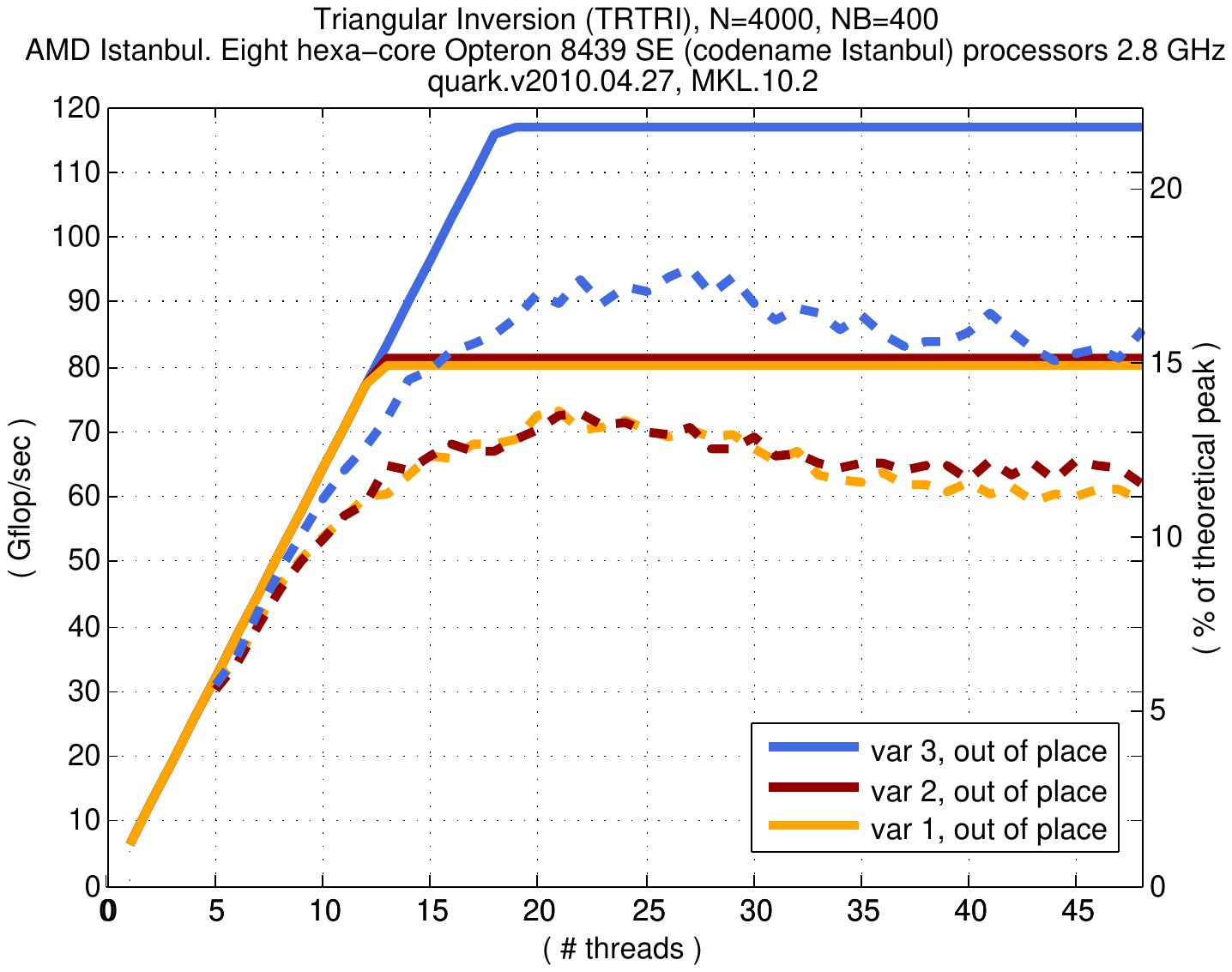}
      }
  \end{tabular}
  \caption{Performance comparison of TILE\_TRTRI for in-place and out-of-place.
Dashed curves represent experimental data, plain curves represent the upper bounds derived in Section 6.
See Table 4 for the critical path lengths of these variants.
}
  \label{fig:TRTRI_threads}
\end{figure}

In Figure~\ref{fig:TRTRI_threads}, the performance of three variants
for TILE\_TRTRI are compared keeping the problem size and tile size fixed while
increasing the number of threads.  Variant 3 outperforms the other two which is
in keeping with the analysis in Section~\ref{sec:TRTRI} where the length of the
critical path for Variant 3 is shorter than that of the others.  Moreover,
Variant 2 outperforms Variant 1 as was the case with the critical path lengths.  
Also note that our upper bounds on performance obtained in Section~6
(plain curves) are reasonably tight.

Considering that the out-of-place variants did introduce some added overhead
due to the necessity of the buffers, of note is the performance gains seen in
Figure~\ref{fig:TRTRI_threads_outofplace} for Variant 1 of TILE\_TRTRI as
compared to the decrease in performance of the other two variants.  In
Table~\ref{tab:cpTRTRI}, it is seen that the added buffers did not shorten the
critical path for Variants 2 or 3, but did improve the critical path for
Variant 1 as is reflected in the numerical experiments.  

\begin{figure}
    \centering
    \includegraphics[width=0.75\textwidth]{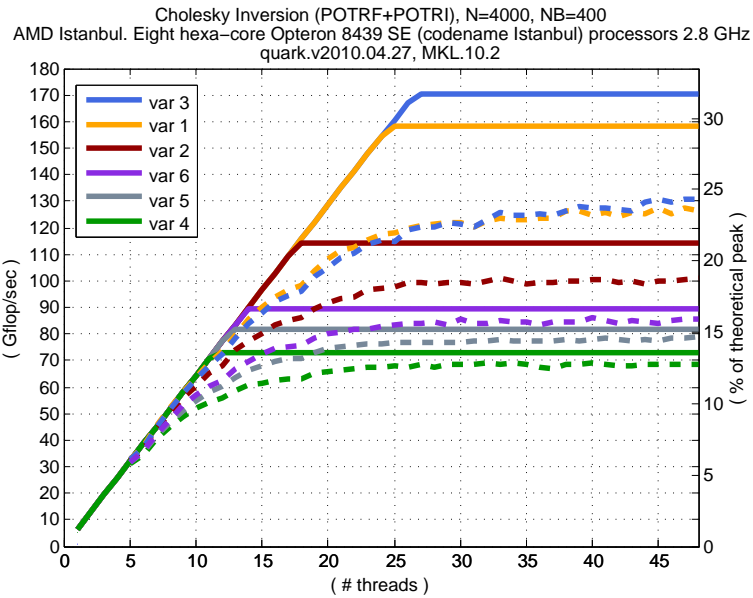}
    \label{fig:CHOLINV_threads}
    \caption{Performance of CHOLINV (in-place). 
Dashed curves represent experimental data, plain curves represent the upper bounds derived in Section 6.
See Table 5 for the critical path lengths of these variants.
}
\end{figure}

Figure~\ref{fig:CHOLINV_threads} provides a comparison of all six variants where
again the matrix size and the tile size are kept constant but the number of
threads are increasing.  This figure clearly mimics the information of
Table~\ref{tab:cpCHOLINV} relative to the number of flops on the critical path
lending credence to the criteria that a better variant has a shorter critical
path.  

\begin{figure}
    \centering
    \includegraphics[width=0.75\textwidth]{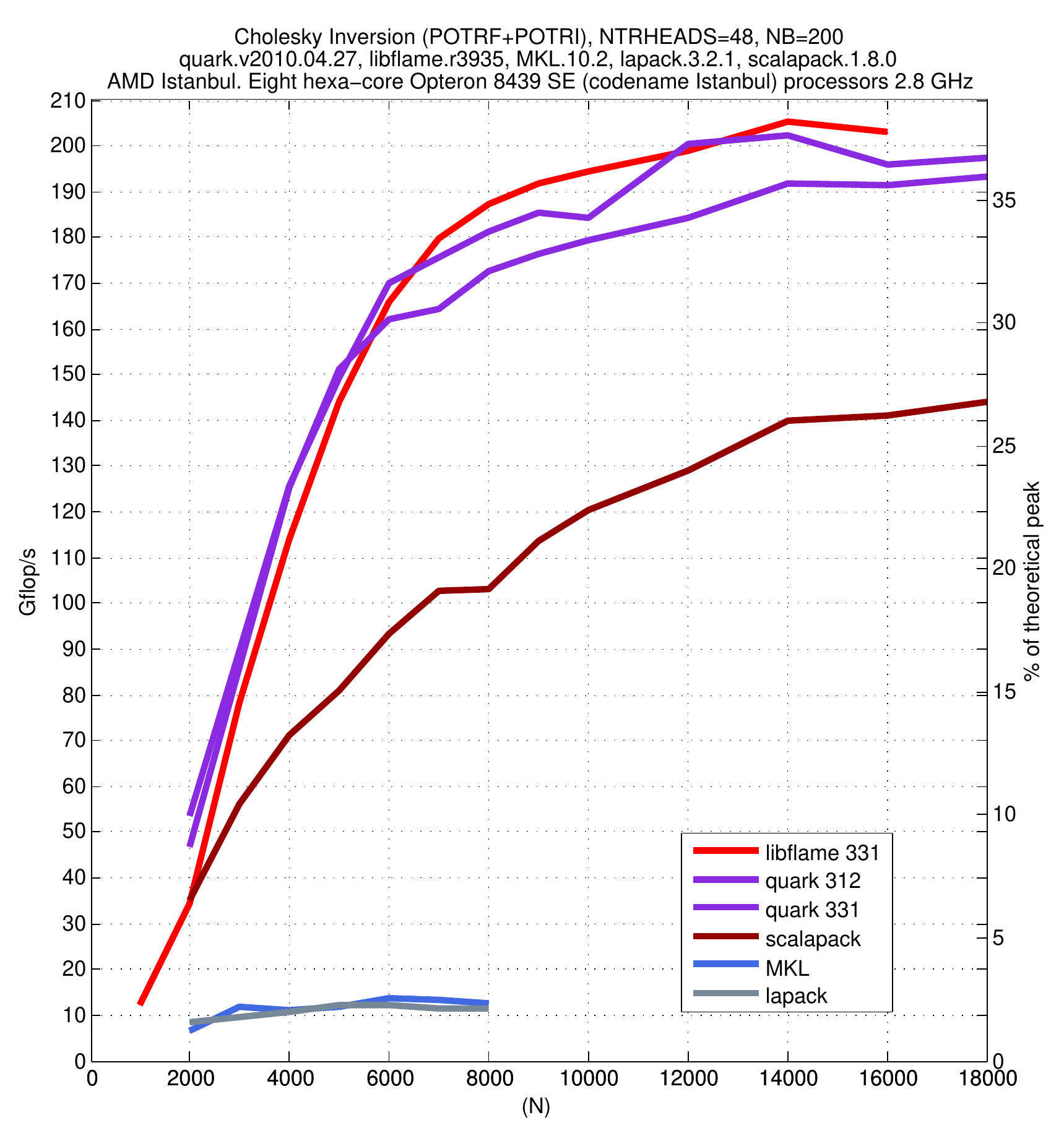}
    \label{fig:CHOLINV_scalability}
    \caption{Performance comparison of LAPACK, MKL, ScaLAPACK, libflame and quark.}
\end{figure}

In order to provide a complete assessment, Figure~\ref{fig:CHOLINV_scalability}
demonstrates a comparison of the complete Cholesky inversion using the dynamic
scheduler quark v2010.04.27 against libflame r3935, MKL v10.2, LAPACK.3.2.1,
and ScaLAPACK v1.8.0.  In this experiment, the number of threads is held
constant at 48, the tile size remains 200, and the matrix size varies.  Once
again, Variant 3 (quark331) shows improvement over Variant 1 (quark312). 
Variant 3 has the shortest flops-based critical path length.

\section{Conclusion}

This paper continues our research on an effective implementation of tiled
Cholesky inversion on multicore platforms~\cite{vecpar10}. Previous
research~\cite{Quintana:2009,BientinesiGunterVanDeGeijn:08} presented
algorithms and performances. In this manuscript, we explain that different
algorithmic variants of the Cholesky inversion algorithm have different
critical path lengths. We provide critical path lengths in terms of the number
of tasks and in terms of the number of flops for all known variants, in place
and out of place. This enables us to understand the scalability of each
variant.

With the current trend in architecture towards multicore, the perspective of
previous algorithms with a focus on the number of flops is now an antiquated
metric.  As more processors are made available, the length of the critical path
becomes the limiting factor and less attention is spent on the total number of
flops.  Our intent is to introduce the length of the critical path as a better
metric for an algorithm.

With this metric we understand why out of the six variants possible for
TILE\_TRTRI, Variant 3 is the most appropriate in the context of Cholesky
inversion: Variant 3 is the one that provides the shortest flops-based critical
path length in this context.

We validate the usefulness of our results with our software on a 48-core
machine and present experimental comparison with LAPACK, ScaLAPACK, MKL, and
libflame. We note that the Cholesky inversion software from this article will
be released in the PLASMA release for SC 2010.

This manuscript focus on parallelism only and neglects (intentionally) any data
transfer issues. This is the reason why the granularity of the problem has
been kept constant all along.  A better understanding of the performance of our
algorithms needs to take into account a data transfer model. This is in our future
work.

When there are few processors or when there is a large number of processors,
our experimental data is often tight with our upper bound on performance. In
between, the discrepancy between our upper bound and the experimental data
can be larger and our future work also aims at reducing this discrepancy.

\section*{Acknowledgement}
The authors thank Ernie Chan and Robert van de Geijn for making the
benchmarking and tuning of libflame as easy as possible. The authors thank Jack
Dongarra for providing access to {\tt ig.eecs.utk.edu} on which the experiments
were performed.

\bibliographystyle{elsarticle-num}
\bibliography{biblio.bib}

\end{document}